\documentclass[aps,prc,preprint,showpacs,showkeys,tightenlines]{revtex4}
\usepackage{epsfig}

\begin{document}

\title{Transition rates and nuclear structure changes in mirror
nuclei $^{47}$Cr and $^{47}$V}

\author{D. Tonev,$^{1,2}$ P. Petkov,$^{1,3}$ A. Dewald,$^{1}$ T.
Klug,$^{1}$ P. von Brentano,$^{1}$ W. Andrejtscheff,$^3$ S. M. Lenzi,$^{4}$
D. R. Napoli,$^{5}$ N. Marginean,$^{5}$ F. Brandolini,$^{4}$ C. A. Ur,$^{4}$
M. Axiotis,$^{5}$ P. G. Bizzeti,$^{6}$ and A. Bizzeti-Sona$^{6}$}

\affiliation{$^1$ Institut f\"ur Kernphysik der Universit\"at zu K\"oln,
D-50937 K\"oln, Germany\\
$^2$ Faculty of Physics, University of Sofia, 1164 Sofia, Bulgaria\\
$^3$ Bulgarian Academy of Sciences, Institute for Nuclear Research and Nuclear Energy, 1784 Sofia, Bulgaria\\
$^4$ Dipartimento di Fisica and INFN, Sezione di Padova, 
Padova,  Italy\\
$^5$ INFN, Laboratori Nazionali di Legnaro, Legnaro, Italy\\
$^6$ Dipartimento di Fisica and INFN, Firenze, Italy}

\date{\today}

\begin{abstract}
Lifetime measurements in the mirror nuclei  $^{47}$Cr and $^{47}$V were
performed by means of the Doppler-shift attenuation method using
the multidetector array EUROBALL, in conjunction with the ancillary detectors
 ISIS and the Neutron Wall. The determined 
transition strengths in the yrast cascades are well described by full 
$pf$ shell model calculations.
\end{abstract}





\maketitle

\section{Introduction}
The investigation of mirror nuclei along the $N$=$Z$ line is of 
considerable interest since it addresses directly the charge 
symmetry of the nuclear forces and the role of the Coulomb effects on 
nuclear structure. The nuclei $^{47}$Cr and $^{47}$V are among 
the heaviest  mirror nuclei where high spin spectroscopy  studies are 
presently accessible without employing
radioactive beams. So far,
they have been the subject of numerous experimental
and theoretical studies. One impressive result is
the establishment~\cite{Bentley-98,Cameron-94} 
of the yrast cascades until the band termination state
(with spin/parity $I^\pi$=31/2$^-$ $\hbar$ for these $1f_{7/2}$
nuclei) and the  successful description of these structures by shell model
(SM)
calculations~\cite{Pinedo-97}. The comparison of the experimental 
excitation energies $E_x$ in the
yrast cascades of $^{47}$Cr and $^{47}$V shown in Fig.~\ref{level-scheme} 
yields the Coulomb energy differences (CED).
Recently, the behavior of the CED [$E_x(^{47}Cr)-
E_x(^{47}V)$] as a function of spin was presented as an evidence for
nuclear structure (alignment of particles) and shape changes 
~\cite{Bentley-98, Bentley2-98, Cameron-93, Lenzi-01}.
One of the aims of the present work is
to verify if these changes are reflected by the corresponding
$B(E2)$ transition strengths 
in both nuclei. Another topic of interest is the comparison of the $B(E2)$'s
with the predictions of the extensive full $pf$ shell model  
calculations~\cite{Pinedo-97}. As already mentioned,
the SM calculations provide a  good description of the
level energies in the investigated nuclei. The preliminary data 
available~\cite{Cameron-98,Brandolini} for the subpicosecond 
lifetimes of interest in
$^{47}$V determined the choice of our experimental tool 
which is the Doppler-shift attenuation method (DSAM).  

\section{Experimental details}
Excited states in $^{47}$Cr and $^{47}$V were populated using the
$2\alpha n$ and $2\alpha p$ exit channels, respectively,  of the
reaction $^{28}$Si + $^{28}$Si. The beam, with an energy of 110 MeV, 
was delivered by the XTU Tandem accelerator of the Laboratori Nazionali 
di Legnaro. The target consisted of 0.85 $mg/cm^2$ Si
(enriched to 99.9 \% in  $^{28}$Si) evaporated on a 
15 $mg/cm^2$ Au backing. 
The deexciting $\gamma$ rays were registered with the EUROBALL III array
~\cite{Euroball III},
using only cluster and clover Ge detectors for the present
measurement.  
Charged particles were detected with the ISIS silicon ball (40
$\Delta E$  - $E$ telescopes)~\cite{Farnea-98}
 and neutrons with the Neutron Wall (50 large volume
liquid scintillator detectors)~\cite{Neutron Wall}. 
Events were collected when at least two $\gamma$
rays in the Ge 
cluster or clover segments and one neutron identified in the Neutron 
Wall were in coincidence or at least three Ge detectors plus one
signal in neutron detector fired in coincidence. Gain matching and efficiency 
calibration of the Ge detectors were performed using $^{152}$Eu and $^{56}$Co 
radioactive sources. A standard add-back correction for Compton
scattering was applied. The data were sorted into coincidence $\gamma$
- $\gamma$ matrices whereby the detection of charged particles and/or
neutrons was required. Under the trigger conditions employed, the  yield of
$^{47}$V was estimated to be 8 \% of the total cross-section while 
that of $^{47}$Cr was found to be about 0.25 \% only.
Therefore the particle gates had to be used to select the reaction channels
of interest out of equally strong or stronger competing exit channels.

For the purposes of a DSAM measurement, the information on the
detection angles of the $\gamma$ rays is of primordial importance because
the Doppler shifts of their energies depend on the angle between the
direction of the recoil velocity and the direction of observation. Therefore 
the cluster and clover detectors of EUROBALL were grouped into 
rings corresponding to approximately the same polar angle with respect to the 
beam axis. Taking into account the segmented structure of these detectors 
leads to the formation of ten rings positioned  at angles of about 72,
81, 99, 107, 122, 129, 137, 145,  155,  and 163$^\circ$. 
In the case of $^{47}$V, the statistics allowed to use (particle gated) 
$\gamma$-$\gamma$ 
matrices where the angular information was conserved on both axes of
the matrix, i.e., the events represented the registration of 
two coincident $\gamma$ rays by the detectors of two particular rings.
In principle, 100 matrices of that type could be constructed corresponding
to all possible two-ring combinations. 
We have sorted 30 such matrices selected on the basis of the quality of the
information provided by them.
Because of the weak reaction channel leading to $^{47}$Cr, in this case 
only ten different matrices were sorted where one of the axes was 
associated with a specific detection angle while on the other axis 
every detector (ring) firing in coincidence was allowed.  

\section{Data analysis}
For the Monte-Carlo (MC) simulation of the slowing-down histories of the 
recoils we used a modified~\cite{Petkov-98,Petkov-99-A} version of
the program DESASTOP~\cite{Winter-83} written by G.Winter. 
In this version, the time-evolution of the recoil velocities  in the target 
and stopper is followed in three dimensions.
The electronic stopping powers used were obtained from the 
Northcliffe and Schilling tables~\cite{Northcliffe-70}
with corrections for the atomic structure 
of the medium along the lines discussed in ref.~\cite{Ziegler-85-1}.
As suggested in ref.~\cite{Keinonen-84}, an empirical reduction 
of $f_n$ = 0.7 was applied to downscale the nuclear stopping power predicted
by the theory of Lindhard, Scharff and Schi\o tt ~\cite{LSS}.
According to the calculations performed, the mean velocity of the recoils
when leaving 
the target was about 3.9 \% of the velocity of light, and they needed
in average 1.2 $ps$ to come to rest. 
The evaporation of charged particles, which is of importance for the
velocity distribution of light residual nuclei, was taken into account in 
the MC simulation and led to better fits of the spectra.
The database of about 10000 
velocity histories was additionally randomized with respect to the 
experimental setup by taking into account the exact position of the 
EUROBALL detectors and their finite size. Complementary details on our 
approach for Monte Carlo simulation can be found in 
refs.~\cite{Petkov-98,Petkov-99-A,Tonev-01}.

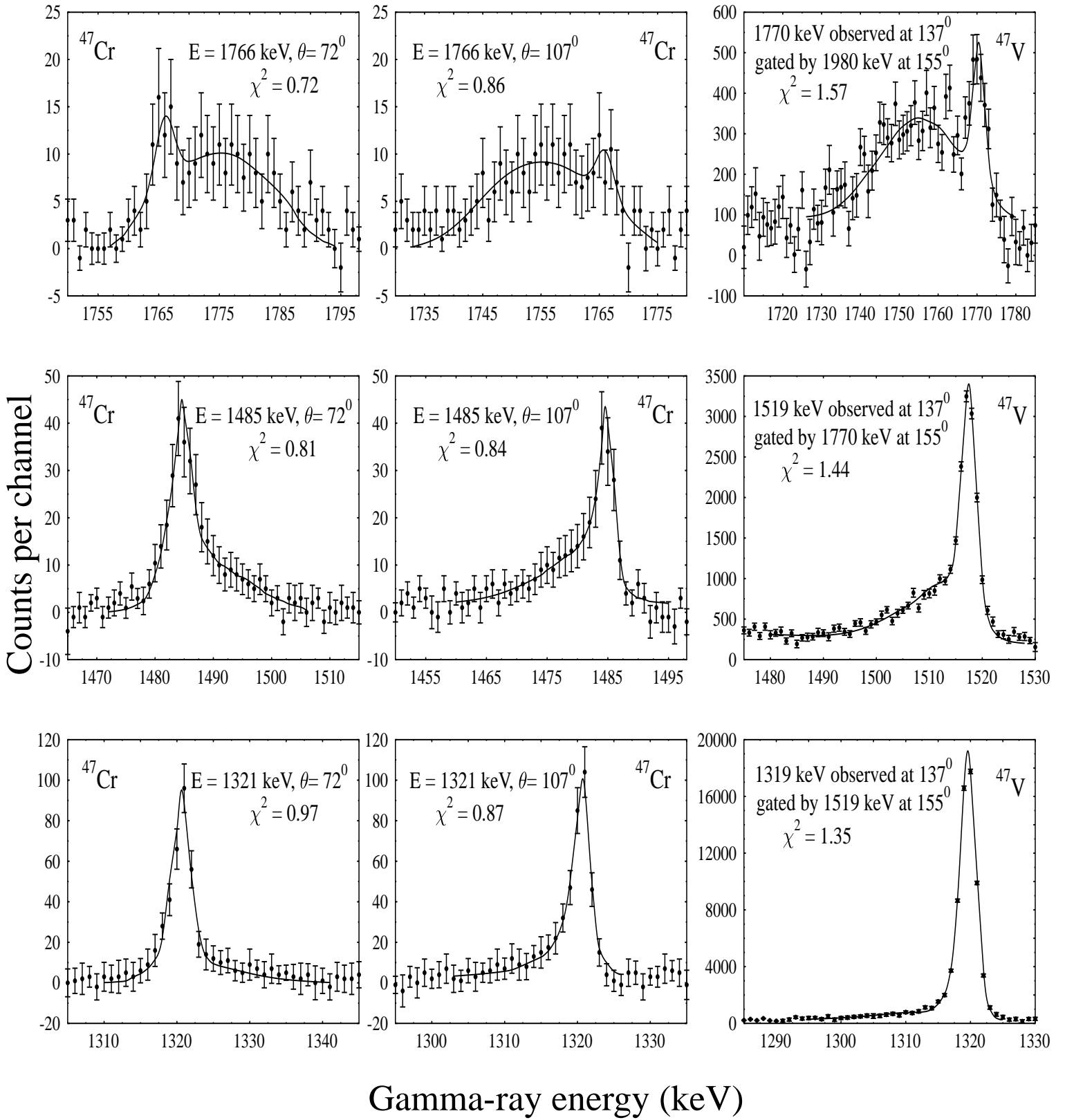
\begin{figure}[p]
\caption{\label{fits} Fits of line shapes of transitions in 
$^{47}$Cr and $^{47}$V measured at different angles.}
\end{figure}

The differences in the acquired statistics for the two mirror nuclei and 
the nature of the sorted
data were the reason for the application of two different approaches for 
data analysis. The 
strength of the $^{47}$V reaction channel made it possible to apply
the newly developed procedure for analysis of coincidence DSAM data
where the gate is set on the shifted portion of the line shape of a transition
directly feeding the level of interest~\cite{Petkov-99-B}. Within this
approach, the timing quality of the gated line shape is improved compared
to the case where the gate includes also fully stopped events because 
they do not bring lifetime information. Moreover, gating from above 
allows the elimination of the uncertainties related to the unobserved 
feeding of the level 
of interest which perturb singles measurements and
coincidence measurements where the gate is set on a transition 
deexciting a level fed by the level of interest. We had to use the 
latter approach for
$^{47}$Cr because of the weakness of its
yield. The details of the analysing procedure used are
presented in refs. ~\cite{Petkov-98,Boehm-93}. 
We only mention here that within its framework it
is possible to investigate the influence of the unobserved feeding
on the derived lifetimes. Fits of line shapes obtained using both approaches
are presented in Fig.~\ref{fits} which illustrates also the quality 
of the data. An error propagation calculation was performed to take into
account the effect of the background subtraction on the errors of the
gated line shapes.

\section{Results} 
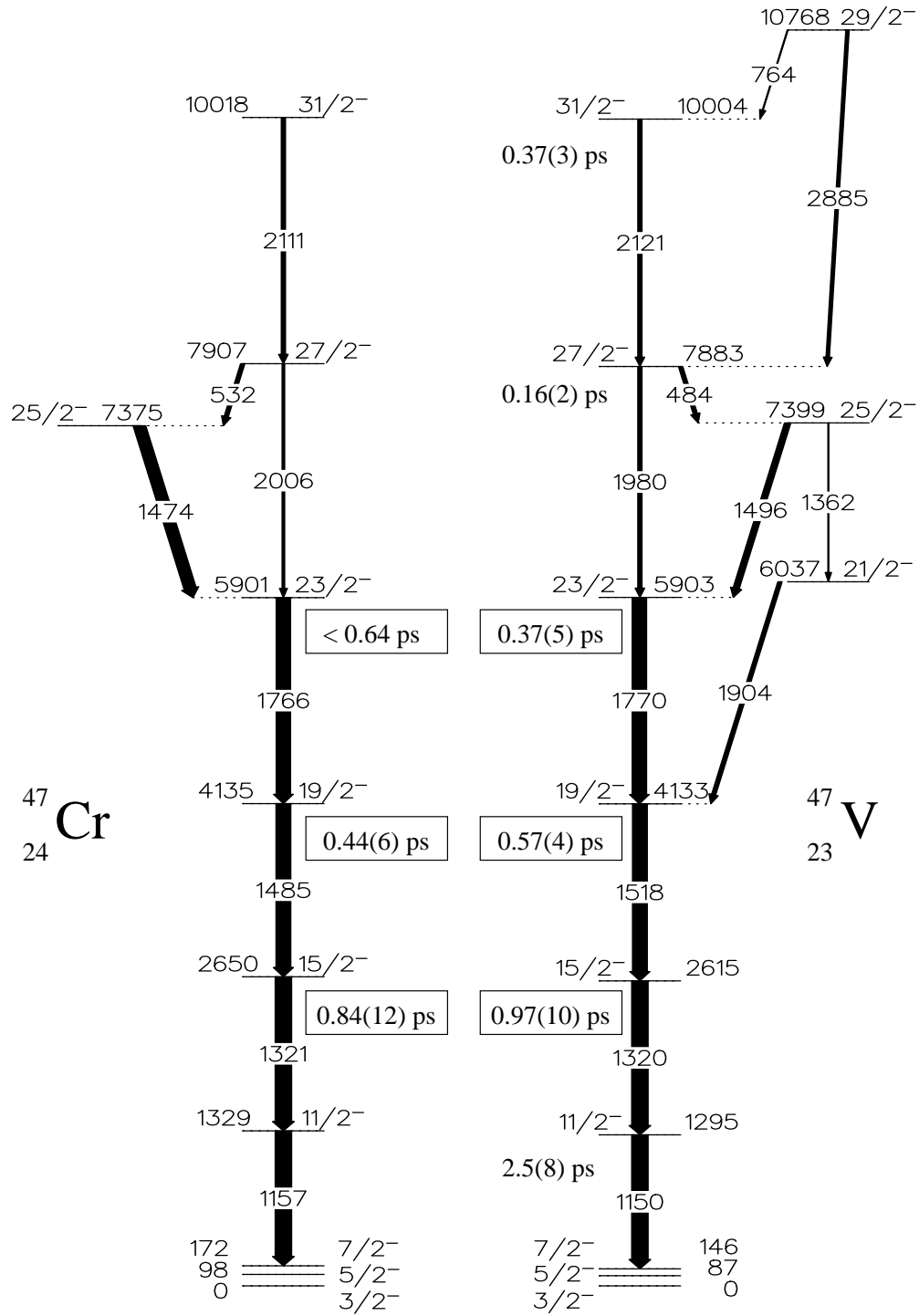
\begin{figure}[p]
\caption{\label{level-scheme} Partial level schemes of  
$^{47}$Cr and $^{47}$V from ref.~\cite{Bentley-98}
 showing mainly the yrast cascades of the two nuclei. Lifetimes
determined in the present work are displayed in framed boxes. Additional
lifetime information from ref.~\cite{Brandolini} is also shown.}
\end{figure}

The results of the analysis are presented in Table~\ref{results}. The
derived lifetimes are indicated also in Fig.~\ref{level-scheme}. 
In the case of $^{47}$V, they
were obtained from the analysis of gated spectra corresponding to 15 different
two-ring combinations. The new lifetime values are in agreement with
the data from earlier measurements. The lifetimes in $^{47}$Cr are 
determined for the first time.  For $^{47}$Cr, the analysed line shapes
where obtained by gating on the 1157 keV transition which does
not show any shifted component. On top of the cascade 
(cf. Fig.~\ref{level-scheme}) the statistics of the line shape
of the 2006  keV transition is too poor and prevents the
precise determination of the lifetime of the 23/2$^-$ level.
In the analysis, we have used the hypothesis that the time behavior of
the unobserved feeding is the same as that of the known feeding.
The intensity balance of the investigated
levels in $^{47}$Cr reveals that the fraction of the unobserved feeding
is less than 10\%. Its influence, studied according to the procedure presented
in ref. ~\cite{Petkov-98}, leads to uncertainties within the statistical
error bars, which makes the lifetime
values derived more reliable.
We estimate the uncertainty due to the imprecise knowledge of the
stopping powers to 10\% and include it in the final errors of the 
lifetimes given in
Table~\ref{results}. It should be  noted that the derivation of lifetimes
in $^{47}$Cr and $^{47}$V in the same experiment makes the determination
of the ratios of the corresponding transition strengths very precise 
since uncertainties related to the stopping powers nearly cancel.
Moreover, our investigation represents the first determination of picosecond
lifetimes of high-spin states in mirror nuclei from the 1$f_{7/2}$ shell 
associated with strong intraband transitions.

\section{Discussion}
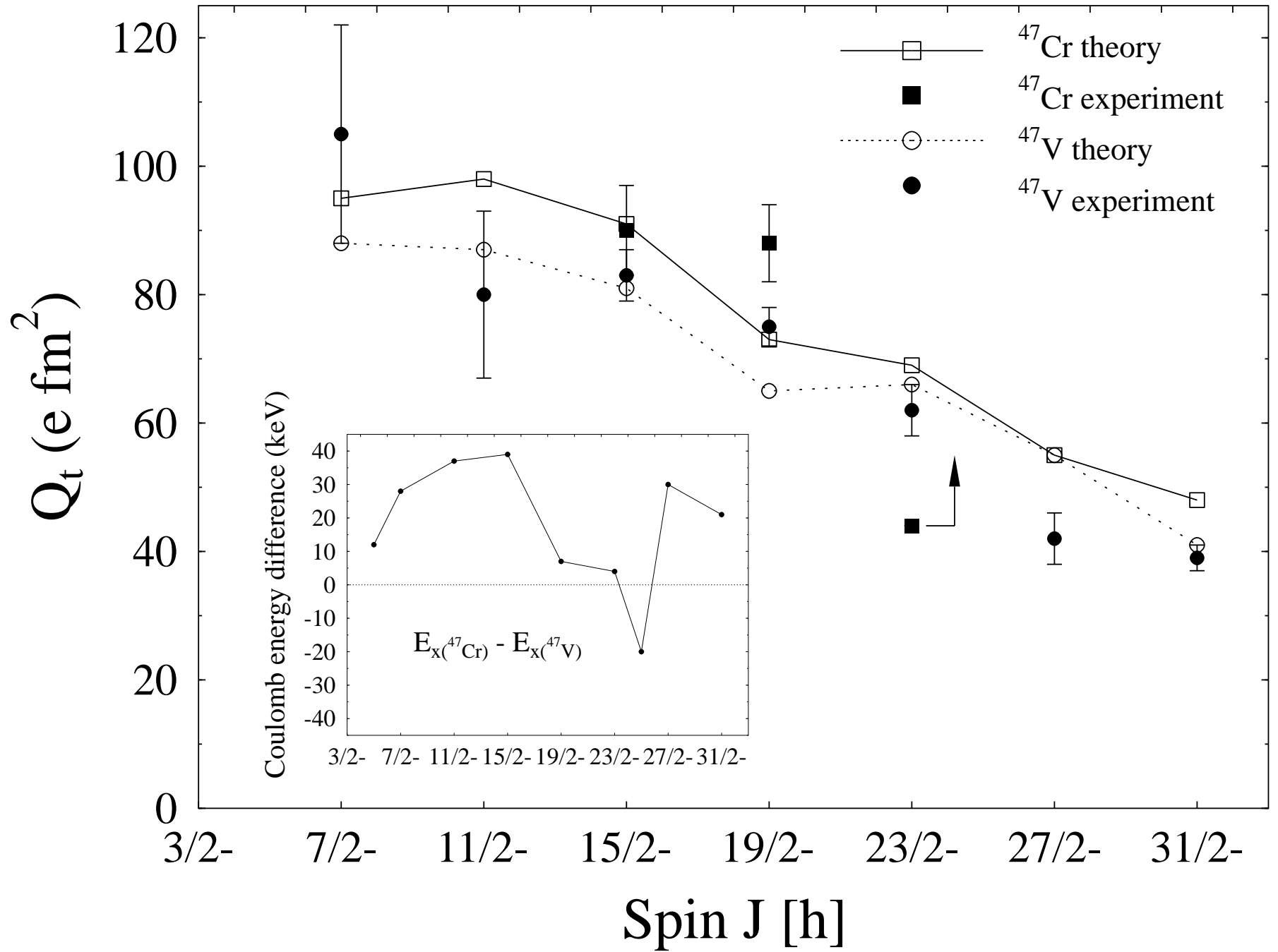
\begin{figure}[p]
\caption{\label{Q_t} Measured and theoretical transition quadrupole
moments. The inset shows the behavior of the CED between 
$^{47}$Cr and $^{47}$V. The arrow indicates that the $Q_t$ value for the
23/2$^-$ level in $^{47}$Cr is only a lower limit.}
\end{figure}

Nuclei in the middle of the 1$f_{7/2}$ shell exhibit rotational 
properties at low and intermediate spins since the number of valence
particles is sufficient to generate such collective effects.
One very interesting result of the SM calculations~\cite{Pinedo-97} for
this spin region in  $^{47}$Cr and $^{47}$V is the closeness of the 
predicted behavior of the $B(E2)$'s to the expectations of the
rotational model~\cite{Bohr-75}.
We have derived from the $B(E2)$ data transition quadrupole
moments $Q_t$ using the equation
\begin{equation}\label{BE2-rigid-rotor}
B(E2,I \rightarrow I-2) = \frac{5}{16\pi}<IK20|I-2K>^2(Q_t)^2
\end{equation}
The $Q_t$ values obtained  in this way correspond to
the intrinsic quadrupole moment $Q_0$ of the rotational model for a band
with $K=3/2$. Using the relation~\cite{Bohr-55}
\begin{equation}\label{Q_0}
Q_0 = \frac{3}{\sqrt{5\pi}}ZeR_0^2\beta(1+ \frac{1}{8}
        \sqrt{\frac{5}{\pi}}\beta)
\end{equation} 
we have also calculated the quadrupole 
deformation $\beta$ which can be associated
with the measured transition strengths under the assumption of an
approximate validity of the rotational model. The experimental $Q_t$ values
are compared to the theoretical~\cite{Pinedo-97} ones in 
Table~\ref{results} where the deduced values of $\beta$ are given too. The spin
dependence of the $Q_t$'s can be seen more easily in Fig.~\ref{Q_t} where
data available for other levels in $^{47}$V (taken from 
ref.~\cite{Brandolini}) and the full pf shell model
predictions ~\cite{Antoine} are displayed. There is a very good agreement 
between the experiment and the theory. This indicates that the effective 
charges used $(e_{\pi}=1.5, e_{\nu}=0.5)$ are appropriate.

The information presented in Fig.~\ref{Q_t} and Table~\ref{results} 
sheds light on several aspects of the nuclear structure of $^{47}$Cr and
$^{47}$V.
The behavior of the $Q_t$'s determined in the
present DSAM measurement in $^{47}$V shows a systematic decrease
with increasing spin. The data for $^{47}$Cr is less conclusive, but they 
exclude an increase of the $Q_t$'s.
The inset to Fig.~\ref{Q_t} shows the
Coulomb energy differences (CED) between the levels of the yrast cascades
of $^{47}$Cr and $^{47}$V. Obviously, our $Q_t$ values characterize
levels at which substantial changes in the CED occur. These changes were
presented by Bentley et al.~\cite{Bentley-98} as an evidence for 
changes in the underlying nuclear 
structure and the nuclear shape. The effect is explained by an
alignment of a pair of 
protons in $^{47}$Cr around spin $I$ = 19/2 $\hbar$ which leads to a 
reduction of their spatial correlations
and thus of the Coulomb energy. This energy is not
affected in $^{47}$V where two neutrons align. The present $Q_{t}$ values
for $^{47}$V point at a sharper decrease of these quantities after the 
$I$ = 19/2 $\hbar$ level. This trend may be associated with the nuclear 
structure changes inferred earlier in Ref.~\cite{Bentley-98} by observation of the 
behavior of the CED. However, the present data for $^{47}$Cr does not allow
to make the same conclusions and this point has to be clarified  in the 
future by
a more precise measurement of the lifetimes of the levels above the 
$I$ = 19/2 $\hbar$ one.
Such a development with increasing spin has to be expected since the valence
space of the 1$f_{7/2}$ nuclei considered is not very large. 
Its capability to generate collective effects gets exhausted when
approaching the band-termination state whose angular momentum is built
only from the contributions of individual nucleons. 
Since all valence protons align in both nuclei $^{47}$Cr and $^{47}$V, the 
CED
close to the band termination state reach similar values to those at the
lowest spins ~\cite{Bentley2-98}. 
Recently, the behavior of the CED in several $f_{7/2}$ mirror nuclei
has also been studied in the framework of the shell model ~\cite{Zuker-last}.
The $A=47$ CED have been reproduced with good accuracy by 
including the change in nuclear radii
along the yrast states together with a renormalization of Coulomb matrix
elements in the pf shell. We note that there is a small disagreement between
the experiment and the shell model calculations at spin $I^\pi$ = 19/2$^-$
(Fig.~\ref{Q_t}), which indicates that the shape changes are not fully
satisfactory described by the model.

In a different
theoretical framework, the cranked Nilsson-Strutinski calculations quoted in 
ref.~\cite{Bentley-98} predict for $^{47}$Cr a gradual change of the
deformation from $\epsilon$ = 0.21 to about $\epsilon$ = -0.03 (oblate)
at the $I^\pi$=31/2$^-$ band-termination states. Our values for $\beta$ at
the beginning of the spin region where the changes of shape start
(cf. Table~\ref{results}) are a bit higher 
($\epsilon$ $\approx$ 0.95$\beta$)
than these predictions. The agreement is even better if we reduce the
deformation derived from the transition probabilities by a factor of about
1.1  to compare it with the deformation of the mean nucleon field 
(see e.g. ref.~\cite{Nazarewicz-90}).
As indicated clearly by the data for $^{47}$V, the deformation decreases 
with increasing spin.  This evolution is a good example for an interplay
between collective and microscopic degrees of freedom.

The nuclear forces are usually considered to be  charge symmetric
and one expects the same structural properties in mirror nuclei with
differences caused only by the Coulomb interaction. The extent of the
mirror symmetry can be already deduced from Fig.~\ref{level-scheme} 
which compares the level schemes of $^{47}$Cr and $^{47}$V. Our
lifetime investigation shows that this symmetry is  
observed also in the transition  probabilities. The small differences 
observed can be understood in  terms of the extra proton in 
$^{47}$Cr. The comparison of the
experimental $Q_t$ values in these mirror nuclei yields information
on the differences in the wave functions of the corresponding states
and the transition quadrupole moments.
To disentangle these two factors is beyond the
scope of the present work but may help further refinement of the
theory. 

\section{Conclusions}
To conclude, DSAM lifetime measurements were carried out with the
multidetector array EUROBALL. The results of the analysis,
partly achieved with a new precise procedure~\cite{Petkov-99-B}, provide
valuable information on the transition strengths in the yrast cascades of the
mirror nuclei  $^{47}$Cr and $^{47}$V. The behavior of the transition
strengths with spin is well described by full $pf$ shell model
calculations. In this way, a test of the isospin
symmetry in mirror nuclei is performed on the basis of the determined
$B(E2)$ values.

\begin{acknowledgments}
Two of us (D.T. and P.P.) are grateful for the kind hospitality of the 
Cologne University. This work was funded by the BMBF under contracts 
No. 06 OK 862 I (0) and 06 OK 958 and by the European Commission through 
the Contract ERBFMCT980110 E. U. TMR Programme. 
The support of the Bulgarian National Research 
Foundation (BNRF) under contract Ph.801 is also appreciated.
\end{acknowledgments}

\newpage

\newpage
\clearpage
\small
\begin{table}
\caption{\label{results} Analysed transitions, derived lifetimes $\tau$,
and transition quadrupole moments $Q_t$. The deformation $\beta$ determined
using eq.\ref{Q_0} is displayed in the last column. 
Branching ratios necessary for the derivation of $Q_t$ were taken from 
ref.~\cite{Cameron-94}. 
Uncertainties of the experimental
quantities are shown in brackets. \hfill}
\end{table}
\begin{ruledtabular}
\begin{tabular}{ccccccc}
Transition & E$_\gamma$ & $\tau$ & $\tau_{previous}$ &
   $Q_t^{experiment}$   & $Q_t^{theory}$ & $\beta$ \\
           & $keV$      & $[ps]$ & $[ps]$            &
      $e fm^2$          &  $e fm^2$      &         \\
\hline
           &            &           & $^{47}$Cr &          &       &      \\
15/2$^-$ $\rightarrow$ 11/2$^-$ & 1321 & 0.84 (12) &   -       &
       90 (7)           &    91     &  0.255 (18)    \\
19/2$^-$ $\rightarrow$ 15/2$^-$ & 1485 & 0.44 (6) &   -       &
       88 (6)           &    73     &  0.248 (16)    \\
23/2$^-$ $\rightarrow$ 19/2$^-$ & 1766 & $<$ 0.64  &   -       &
     $>$ 44             &    69     &  $>$ 0.127     \\
\hline
           &            &           & $^{47}$V &          &       &      \\
15/2$^-$ $\rightarrow$ 11/2$^-$ & 1320 & 0.97 (10) & 
                  0.99 (10) ref.\cite{Brandolini}       &
       83 (4)           &    81     &  0.246 (12)    \\
                              &        &           & 
                  $>$ 2.5 ref.\cite{Cameron-98}              &
                        &           &                 \\
19/2$^-$ $\rightarrow$ 15/2$^-$ & 1518 & 0.57 (6) &          
                  0.60 (6) ref.\cite{Brandolini}        &
       75 (3)           &    65     &  0.221 (8)    \\
                              &        &           & 
                 0.70 (10) ref.\cite{Cameron-98}              &
                        &           &                 \\
23/2$^-$ $\rightarrow$ 19/2$^-$ & 1770 & 0.37 (5) &   
                  0.35 (3) ref.\cite{Brandolini}        &
       62 (4)           &    66     &  0.185 (12)    \\
                              &        &           & 
                 0.50 (7) ref.\cite{Cameron-98}              &
                        &           &                 \\
\end{tabular}
\end{ruledtabular}

\end{document}